\documentclass{aa}  

\usepackage{graphicx}
\usepackage{subfigure}
\usepackage{txfonts}
\usepackage{siunitx}
\usepackage{booktabs}
\usepackage{blindtext}
\usepackage{hyperref}

\begin{document}

  \title{New constraints on the average escape fraction of Lyman continuum radiation in $z\sim 4$ galaxies from the VIMOS Ultra Deep Survey (VUDS)
                             \thanks{Based on data obtained with the European 
                                       Southern Observatory Very Large Telescope, Paranal, Chile, under Large
                                       Program 185.A--0791. }
                  }

           \author{F. ~Marchi\inst{1}
                   \and L.~Pentericci\inst{1}
                   \and L.~Guaita\inst{1}
                   \and B.~Ribeiro\inst{2}
                   \and M.~Castellano\inst{1}
                   \and D.~Schaerer\inst{3,7}
                   \and N.P.~Hathi\inst{2,11}
                   \and B.~C.~Lemaux \inst{10}
                   \and A.~Grazian\inst{1}
                   \and O.~Le F\`evre\inst{2}
                        \and B.~Garilli\inst{6}
                        \and D.~Maccagni\inst{6}
                        \and R.~Amorin\inst{8,9}
                        \and S.~Bardelli\inst{4}
                        \and P.~Cassata\inst{5}
                        \and A.~Fontana\inst{1}
                        \and A.~ M. ~Koekemoer\inst{11}
                        \and V.~Le Brun\inst{2}
                        \and L. A. M.~Tasca\inst{2}
                        \and R.~Thomas\inst{2}
                        \and E.~Vanzella\inst{4}
                        \and G.~Zamorani \inst{4}
                        \and E.~Zucca\inst{4}
                }       
           \institute{INAF–Osservatorio Astronomico di Roma, via Frascati 33, 00040 Monte Porzio Catone, Italy\\
                      \email{francesca.marchi@oa-roma.inaf.it}
                 \and
                         Aix Marseille Universit\'e, CNRS, LAM (Laboratoire d'Astrophysique de
                         Marseille) UMR 7326, 13388, Marseille, France
                 \and
                         Geneva Observatory, University of Geneva, ch. des Maillettes 51, CH-1290 Versoix,
                         Switzerland
                 \and
                         INAF--Osservatorio Astronomico di Bologna, via Ranzani, 1 - 40127, Bologna, Italy
                 \and
                         Instituto de Fisica y Astronom\'ia, Facultad de Ciencias, Universidad de
                         Valpara\'iso, Gran Breta$\rm{\tilde{n}}$a 1111, Playa Ancha, Valpara\'iso Chile
                 \and
                         INAF--IASF Milano, via Bassini 15, I--20133, Milano, Italy
                 \and
                         Institut de Recherche en Astrophysique et Plan\'etologie - IRAP, CNRS, Universit\'e
                         de Toulouse, UPS-OMP, 14, avenue E. Belin, F31400
                         Toulouse, France            
                     \and 
                             Cavendish Laboratory, University of Cambridge,
                               19 JJ Thomson Avenue, Cambridge, CB3 0HE, UK    
                         \and
                                  Kavli Institute for Cosmology, University of Cambridge,
                                         Madingley Road, Cambridge CB3 0HA, UK
                         \and
                                 Department of Physics, University of California, Davis, One Shields 
                                 Ave., Davis, CA 95616, USA
                         \and
                                 Space Telescope Science Institute, 3700 San Martin Drive, Baltimore, MD 21218, USA
                        }
        
           \date{Received; accepted}

   \date{}
        \abstract
        {Determining the average fraction of Lyman continuum (LyC) photons escaping high redshift galaxies is essential for understanding how reionization proceeded in the z>6  Universe. }
        {We want to measure the LyC signal from a sample of sources in the Chandra Deep Field South (CDFS) and COSMOS fields for which ultra-deep VIMOS spectroscopy as well as multi-wavelength Hubble Space Telescope (HST) imaging are available.}
        {We select a sample of 46 galaxies at $z\sim 4$ from the VIMOS Ultra Deep Survey (VUDS) database, such that the VUDS spectra contain the LyC part of the spectra, that is, the rest-frame range $880-910\AA$. Taking advantage  of the HST imaging, we apply a careful cleaning procedure and reject all the sources showing nearby clumps  with different colours, that could potentially be lower-redshift interlopers. After this procedure, the sample is reduced to 33 galaxies. We measure the ratio between ionizing flux (LyC at $895\AA$) and non-ionizing emission (at $\sim 1500 \AA$) for all individual sources. We also produce a normalized stacked spectrum of all sources. }
        {Assuming an intrinsic average $L_{\nu}(1470)/L_{\nu}(895)$ of 3, we estimate the individual and average relative escape fraction. We do not detect ionizing radiation from any individual source, although we
        identify a possible LyC emitter with very high Ly$\alpha$ equivalent width (EW).  From the stacked spectrum  and assuming a mean transmissivity for the sample,  we measure a relative  escape fraction $f_{esc}^{rel}=0.09\pm0.04$. We also look for correlations between the limits in the LyC flux and source properties and  find a tentative correlation between LyC flux and the EW of the Ly$\alpha$ emission line.}
        {Our results imply that the LyC flux emitted by  $V=25-26$ star-forming galaxies at z$\sim$4  is at most very modest, in agreement with previous upper limits from studies based on broad and narrow band imaging.} 
        \titlerunning{New constraints on LyC emission}
        \authorrunning{Marchi F.}
   \maketitle
        
        \section{Introduction}
        Studying the process of  hydrogen reionization 
        in the early Universe and identifying the sources of UV radiation responsible for this process are amongst the most challenging tasks of modern extra-galactic  astronomy.
        The radiation able to ionize the neutral hydrogen falls at wavelengths shorter than $912\, \AA$ (Lyman continuum, LyC) and is produced by massive OB-type stars in young
        star clusters and by active galactic nuclei (AGN).  Therefore, star-forming galaxies and AGN  at $z\sim 6-7$  are most likely responsible for this phenomenon \citep{robertson15, giallongo15}. However, it is extremely difficult to directly measure  their contribution since at  redshift  higher than $z=5,$ the intergalactic medium (IGM) becomes completely  opaque to LyC photons and prevents the direct detection of Lyman continuum flux \citep{inoue:transmissivity,worseck14}. In some cases it could be possible to detect LyC emission at higher redshifts if the object is very bright in LyC and  resides in a particular line of sight that favours the escape of LyC photons. These are, however, particular cases that are not able to give us statistics on the populations of LyC emitters at high redshifts.
        
         We therefore rely on the study of $z< 4.5$ galaxies to understand the physical properties of the  objects emitting LyC radiation and later infer if these properties are more common during the reionization epoch. The LyC radiation from galaxies between redshifts of 2.5 and 4.5  can, in principle,  be detected from the ground, while  for galaxies at lower redshift, we must rely on space-based observations. The ionizing radiation is attenuated by neutral gas and dust in the interstellar and circumgalactic medium of the sources. Therefore, the detection of LyC emission in individual galaxies is a rather difficult task. 
         
        There are  different methods to evaluate the \emph{LyC escape fraction}, that is,
        the fraction of HI-ionizing photons that are not
        absorbed by the interstellar medium (ISM) and are thus free to
        ionize the neutral hydrogen in the IGM.
        The first one relies on narrow-band photometry, that is, the study of  galaxies in a given (narrow)  redshift range such that  the supposed LyC flux would fall precisely in a  narrow band filter.
        With this method, the narrow band collects precisely  the flux in the  LyC range: however, because of the limited redshift range probed by a typical narrow band filter, only a low number of sources can be studied at the same time. \cite{lucia} used this technique to analyse a sample of 67 galaxies at $z\sim 3$ (some also from VUDS), finding no individual detections and an upper limit on the relative escape fraction of ionizing photons of $12\%$ for the entire sample. \cite{mostardi15} also applied the narrow band technique to a sample of 16 $z\sim 3$ Lyman-break-selected star-forming galaxies (LBG) and narrow-band-selected Lyman-$\alpha$ emitters, finding one LyC emitter candidate among the LBGs at $z=3.14$ with a relative escape fraction of $f_{esc}^{rel}\sim 75\%-100\%$. Finally, earlier studies by \cite{iwata09} identified several candidate emitters amongst the LAE (125 galaxies) and LBG (73 galaxies) populations of the SSA22 protocluster region at $z=3.09$. 
        
                An alternative  method consists of using broad-band imaging to capture the LyC flux. In this way, one can simultaneously study a much larger number of galaxies in a broader redshift range.
         The drawback of this technique resides in the fact that the filter contains both LyC and non-LyC flux from the source and therefore a more  careful analysis  is needed to disentangle the two components. \cite{grazian16} and \cite{boutsia} applied this technique on  samples (37 and 11 galaxies, respectively) of $z\sim 3$ star-forming galaxies with secure redshifts, finding no significant detection. They used a relatively narrow redshift range so that the U filter on LBC would sample only the LyC radiation, and found average upper limits of $f_{esc}^{rel}\lesssim 2\%$ and $f_{esc}^{rel}\lesssim 5\%$, respectively.
         
         Finally, it is also possible to  directly analyse spectral observations and measure the flux in the LyC region of the spectrum. Fewer galaxies have deep enough spectroscopic observations compared to those with photometric ones. \cite{giallongo02}, \cite{steidel}, \cite{shapley}, \cite{slavo} and \cite{shapley16} used this method to evaluate the escape fraction of high-redshift galaxies finding several  LyC emitter candidates. 
        
        All the above  methods are affected by the problem of line of sight (LoS) contamination that is the main limitation of LyC studies when imaging and spectroscopic observations are taken from the ground \citep{vanzella12}. Indeed low-redshift galaxies can  mimic the LyC emission from high-redshift sources if they are located very close to the target galaxies and the spatial resolution does not allow us to distinguish them. These nearby contaminants can only be identified in
        high-resolution HST images and appear blended in ground-based observations. 
        In most cases,  the putative  LyC emission appears in HST images offset
        with respect to the main
        optical galaxy \citep{nestor,mostardi},  indicating  the presence of a possible lower-redshift contaminant. To date, there are only two galaxies with an unambiguous LyC detection at high redshifts. The first one is the object \emph{Ion2}, a very compact
        star-forming galaxy at z = 3.212, detected by \cite{vanzella15} from deep imaging. It was analysed in detail by \cite{debarros16} and confirmed by \cite{vanzella16}. A second clear Lyman continuum emitter
         was recently identified  at $z=3.15$ by \cite{shapley16}. They used the spectroscopic technique combined with high-resolution HST imaging and inferred a value of $f_{esc}\geq 42\%$ at $95\%$ confidence  for this source.

        In this work,  we   study the LyC emission of a sample of galaxies at $z\sim 4$  using ultra-deep  spectroscopic observations from the VUDS collaboration \citep{lefevreVIMOS}  paying particular attention to minimise the possibility of having nearby lower-redshift contaminants close to our targets. This can be done thanks to the availability of deep multicolour HST imaging.
        The paper is organized as follows: in Sect.\ 2, we describe the sample selection; in Sect.\ 3, we measure the LyC emission in individual spectra and define the quantities needed to evaluate the relative escape fraction; in Sect.\ 4, we present the  stack of the spectra  and in Sect.\ 5, we describe  the results; finally, we discuss the results in view of future spectroscopic  surveys in Sect.\ 6. Throughout the paper we
        adopt the $\Lambda$ cold dark matter ($\Lambda$-CDM) cosmological model ($H_0 = 70 km\, s^{-1} Mpc^{-1}$, $\Omega_M = 0.3$ and $\Omega_{\Lambda} = 0.7$).
        All magnitudes are in the AB system.

        \section{Sample selection}
        Our sample contains  galaxies   selected from the \emph{VIMOS Ultra Deep Survey}
        \citep[VUDS]{lefevreVIMOS} database \footnote{\href{url}{http://cesam.lam.fr/vuds/DR1/}}, which is the largest  spectroscopic survey of galaxies at $z>2$. VUDS   acquired spectra to identify approximately $7000$ galaxies at $2\leq z\leq 6$. We selected all galaxies with reliable redshifts  in the range $3.5 \leq z \leq 4.5$ (see below) and HST
         multi-wavelength coverage. The choice of the redshift interval relies on our aim to measure a possible LyC
        signal in the wavelength range covered by the VUDS spectra ($3800-9400\, \AA$).
We therefore need sources at $z > 3.5$  to have the LyC interval in  the spectrum. 
        The upper redshift limit is instead due to the fact that the Inter Galactic Medium is almost totally opaque at $z > 4.5$ \citep{madau95,inoue:transmissivity}. 
        
        All the galaxies with VUDS reliability flags 3 and 4, corresponding to a probability greater than $95 \%$ for the spectroscopic redshift to be correct, were included in the sample. Only in the cases where the photometric redshift was consistent with the measured spectroscopic redshift,  within  1$\sigma$, we also included those with flags 2 and 9, corresponding in both cases to a probability of approximately $80\%$ for the redshift to be correct with the difference that the redshift of flag 9 objects is based only on one clear emission line in the spectrum \citep[see][for more details]{lefevreVIMOS}. In this procedure, we used CANDELS and MUSYC photometric redshifts \citep{dahalen13,musyc}.
We remark here  the  excellent quality of the  flux calibration of the VUDS spectra that, if not done properly, could significantly affect the evaluation of $f_{esc}^{rel}$. The spectra are calibrated using spectrophotometric standard stars with a relative flux accuracy of better than $\sim$5\% over the wavelength range $3600$ to $9300\AA$. In addition, each spectrum is corrected for atmospheric extinction and for wavelength-dependent slit losses due to atmospheric refraction, taking into account the geometry of each source as projected into the slit \citep{thomas14}. The spectra are also corrected for the galactic extinction \citep{lefevreVIMOS}.

 The last requirement in our sample selection procedure is the HST coverage: observations in at least two HST bands are necessary  to identify possible foreground contaminants that can mimic the LyC emission from the target (see Subsection \ref{subsec:cleaning}). 
        
        In total, we selected 46 galaxies with the above characteristics. Out of
        46, 11 sources are in the COSMOS field and have CANDELS coverage \citep{candels}, whereas the remaining sources are in
        the \emph{Extended Chandra Deep Field South} (ECDFS). Of these, 13 of the galaxies are covered by CANDELS multi-wavelength imaging \citep{candels,grogin11} and 22 fall  outside the CANDELS field but within the area covered by GEMS \citep{gems}. We did not consider the sources in the COSMOS area outside CANDELS because, for them, only one  HST  band (F814W) is available and it  would not be deep enough for our purpose. For the sources that have CANDELS coverage, we used the HST bands ACS/F435W, ACS/F606W and F814W (hereafter referred to as B, V and I) for ECDFS, whereas for COSMOS, we used only V and I. Finally,  for the sources covered by GEMS, we used HST/ACS F606W and HST/ACS F850LP (hereafter
referred to as V and z).

                        We checked the position of our galaxies in the CANDELS fields, in the stellar mass vs. UV
                        absolute magnitude $M_{1400}$ plane, where $M_{1400}$ was evaluated from the V band magnitude.  The masses that we used are the reference masses for CANDELS. They were obtained by \cite{santini15} combining, with a median mass approach, different mass estimates derived with the BC03 templates, a Chabrier IMF and various parametrizations of the SFH with some of the models,  also including nebular emission contribution. Comparing our sample to the general population of  galaxies at $3.5 <z< 4.5$ in the GOODS-South field \citep{grazian15}, we conclude that it is representative of the general population of star-forming galaxies with $M_{UV}\sim -20$ in that redshift range.

        \subsection{Cleaning procedure}
        \label{subsec:cleaning}
        To avoid false LyC detection, we applied a careful cleaning procedure to our sample by removing
        all the galaxies with possible projected contamination on an object-by-object basis. Following \cite{lucia}, we proceeded with three different steps. The first consists of identifying  isolated, single-component sources from those with multiple components with a direct inspection of the HST images; we then assume  that the single-component sources are unlikely to be contaminated by foreground galaxies. The second step consists of generating colour images
        for each multi-component source identified  in the previous step and analysing each sub-component.
        To do this, we used the iraf task IMSTACK to generate colour images in B-V-I bands for the galaxies in
        the area covered by CANDELS in the CDFS field,  V-z images
        for those covered by GEMS and V-I images for the sources in COSMOS. An individual component  or a region  of the galaxy with appreciably different colours (see below) with respect
        to the main galaxy  is likely to be an interloper at a different redshift. Therefore, generating these
        images allowed us to reject these contaminated sources from the sample \citep{vanzella12,siana15}.

        Actually, the situation is slightly more complex, as multiple components (hereafter
        clumps) belonging to the same galaxy can be redder or bluer than the main component of
        the galaxy, depending on the properties of their stellar populations, dust and the presence and the
        strength of emission lines falling within the broad-band filters. The best way to take this into account would be to perform accurate  SED fittings of each individual clump of each multi-component galaxy.  However, this accurate analysis is only possible when many different bands are available, as in the CANDELS fields, but would not be possible, for example, for GEMS or the entire COSMOS. In our previous work \citep{lucia}, after an accurate analysis of a small sample of VUDS galaxies with similar V-band magnitude as our objects,  we showed that simple colour cuts can give results that are entirely comparable to a full SED fitting of individual clumps.
        To reject sources on the basis of the colours, we thus used the following criteria 
        :  $\Delta (B-I)>0.2$ and $\Delta (V-I)>0.2$ for CANDELS and $\Delta (V-z)>0.2$ for GEMS. 
The colours of the clumps  were determined by running Source Extractor (SExtractor, \cite{bertin}) on the sources:
Sextractor configuration parameters for background, source extraction and optimal photometry were fixed in order 
to separate the individual clumps and maximise the signal-to-noise ratio.
        
\begin{figure}
        \centering
        \includegraphics[width=\hsize]{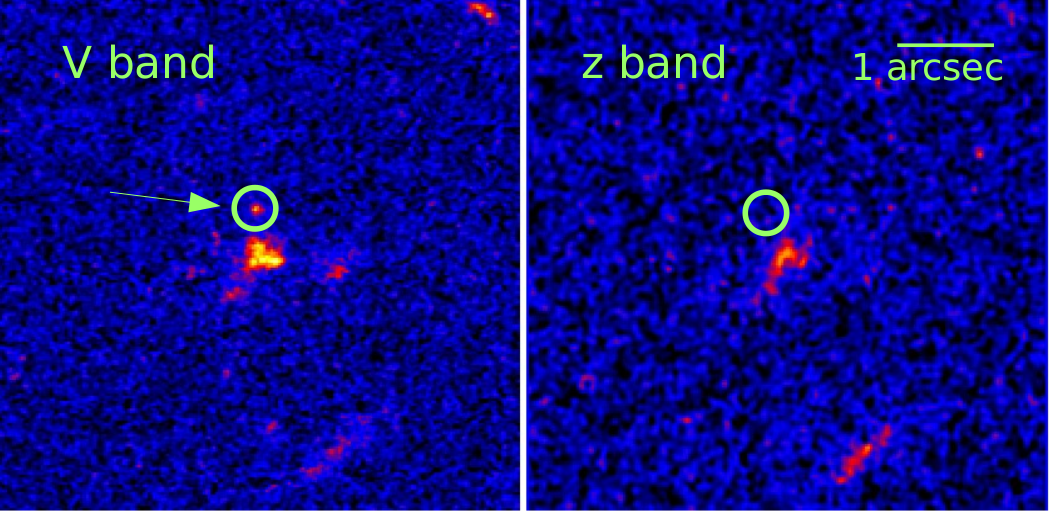}
        \caption{Example of an object discarded in the second step of the cleaning procedure. This is 
         a galaxy at $z=3.5953$ within GEMS coverage. In the figure, we show the two images of the source in the V band and in the z band.
        This was rejected from the sample due to the
        presence of a clump in the V
        band image that is not visible in
        the z band image.}
        \label{fig:530063733_discarded}
        \end{figure}

        In \cite{lucia}, we showed that relying on colours alone is a conservative approach because this selection tends to overestimate the number of contaminated sources. The purity of our sample is, however, our primary concern, since the presence of contaminants is the most worrying aspect in the search for LyC emitters \citep{vanzella12,mostardi15}. With the second step of the cleaning procedure, we excluded eight sources from the sample: one of these is shown in Fig. \ref{fig:530063733_discarded}.
        
         The last step of our cleaning procedure consists of directly inspecting the two-dimensional spectrum of each remaining source   to look for
        the presence of defects in the  spectral region of interest (slit-border defects, sky line residuals etc.)  and to check that the spectra do not show hints of possible contamination. With this direct inspection, we rejected five further sources from the sample. Three of these show slit-border defects in the LyC region of the spectrum, whereas the other two seem to be contaminated by other sources. Both these galaxies were identified in the
        previous step as multi-component but were kept in the sample because they did not show  a large colour difference in  V-z. However, their two-dimensional spectra show an emission in the LyC     region that is shifted in the y-axis compared to the rest, meaning that it is associated only to one component  of the galaxy. Unfortunately both these galaxies are in the GEMS area where we only have two bands: it is therefore harder to determine if the clump really belongs to the galaxy or is at a different redshift. We decided  to  discard the two galaxies from the sample. It is worth noting that all the galaxies rejected in this step of the cleaning procedure are covered by GEMS for which, in the previous step, we could check only one colour. This final step is therefore complementary to the second step of the cleaning procedure for the sources in GEMS.
        \begin{figure}
        \centering
        \includegraphics[width=\hsize]{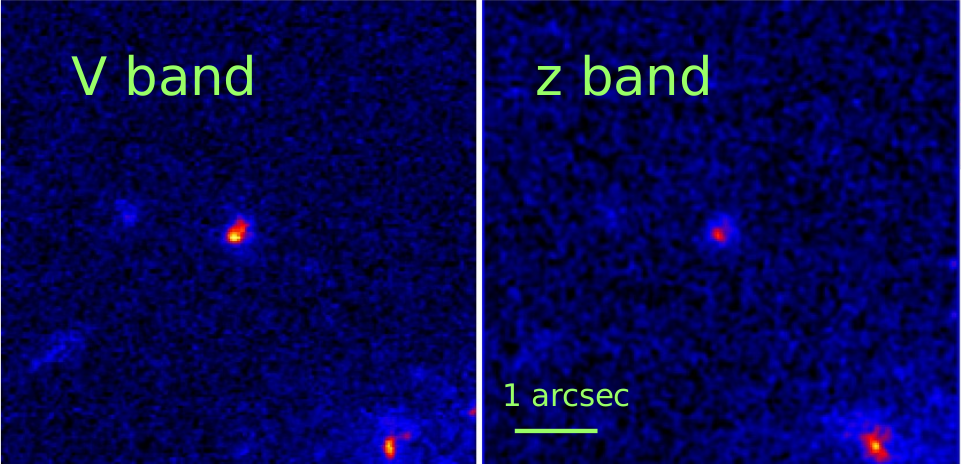} \\
        \includegraphics[width=\hsize]{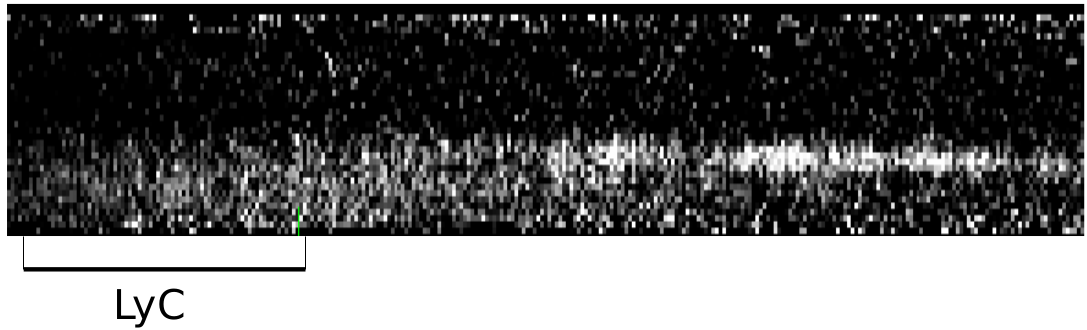}
        \caption{Example of an object
        at $z=3.536$ discarded in the third step of the cleaning procedure. \emph{Upper panel}: images of the source in the V band and in the z band. \emph{Lower panel}: portion of the two-dimensional spectrum of the source that contains the LyC region that is affected by border defects.}
        \label{fig:530063131_discarded}
        \end{figure}
        In Fig. \ref{fig:530063131_discarded}, we show an example of a source that we excluded from the sample due to the presence of slit-border defects.
        
Finally, we remark  that, with this cleaning procedure, it is not possible to identify contaminants that are exactly superposed with the high-redshift galaxy. This is, however, very unlikely and we do not consider this particular case.
        
\begin{figure*}
                \hspace{-0.2cm}
                \includegraphics[width=0.5\textwidth]{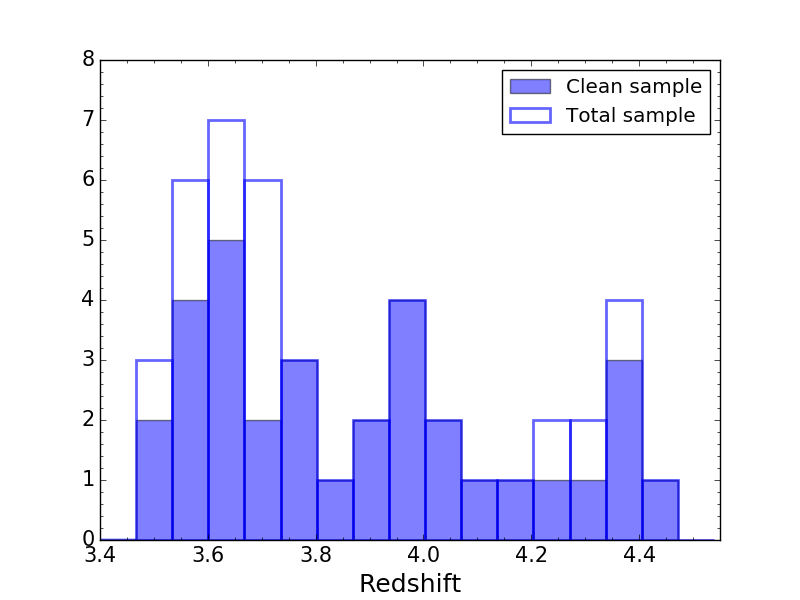}
                \includegraphics[width=0.5\textwidth]{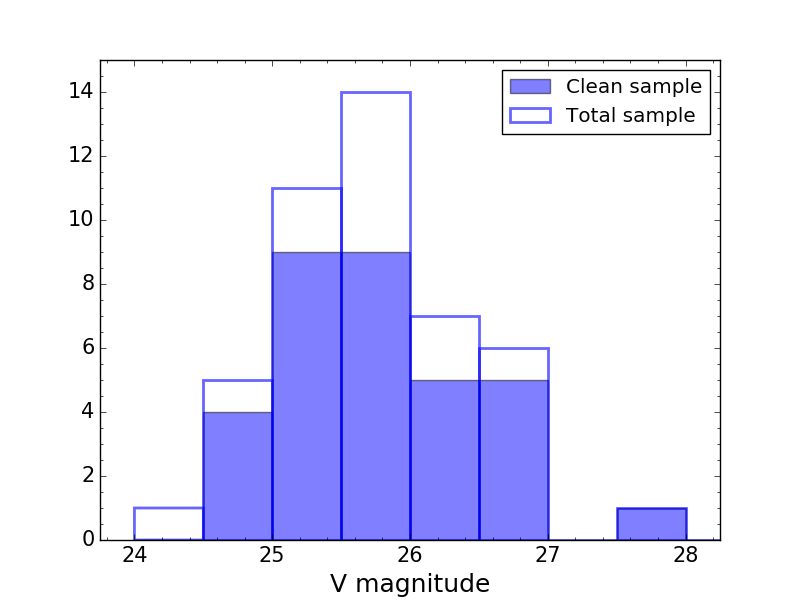}
                \caption{\emph{Left panel}: comparison of the redshift distributions of the total sample (empty
                histogram) and the clean sample (semi-filled histogram). \emph{Right panel}: comparison of the V
                magnitude distributions of the total sample (empty histogram) and the clean sample (semi-filled
                histogram).}
                \label{fig:histo}
                \end{figure*}

        In Table \ref{tab:cleanstep}, we show the number of galaxies in the three HST fields in the different steps of the cleaning
        procedure.      The final sample consists of 33 galaxies and is listed in Table \ref{tab:cleansample}. From here on, we refer to this sample as the clean sample. In Fig. \ref{fig:histo}, we show the redshift and the V magnitude distributions of the galaxies
        in the clean sample (semi-filled histogram) compared to those in the initial sample (empty histogram). From this comparison, it is clear that the excluded sources are a random subset of the entire sample in terms of V band and redshift distribution.

                \begin{table*}
                        \centering
                        \begin{tabular}{l c c c c}
                                \midrule
                                \midrule
                                & COSMOS/CANDELS & ECDFS/CANDELS & ECDFS/GEMS & TOTAL\\
                                \midrule
                                 $\#$initial sources & 11 & 13 & 22 & 46\\
                                 $\#$sources after step 2 & 9 & 11 & 18 & 38\\
                                 $\#$sources after step 3 & 9 & 11 & 13 & 33\\
                                \bottomrule
                                \vspace{0.1cm}
                        \end{tabular} 
                        \caption{Number of sources in the different steps of the cleaning procedure in the three
                        fields. Note that we indicate, with ECDFS/GEMS, the area of GEMS not covered by CANDELS. `Initial sources',  `sources
after step 2' and  `sources after step 3' indicate the number of sources before the cleaning procedure is applied, after the colour cleaning and after the
spectral check, respectively.}
                        \label{tab:cleanstep}
                \end{table*}
        
                \begin{table*}
                \centering
                \scriptsize{
                                \begin{tabular}{l c c c c c c c c c c r}
                                \toprule
                                \toprule
                                (1) &(2) & (3) &(4) & (5) & (6) & (7) & (8) & (9) & (10) & (11) & (12)\\
                                ID$_{VUDS}$ & ID$_{HST}$ & Ra & Dec & z$_{spec}$ & flag & V& $f_{\lambda}(895)$ & err$f_{\lambda}(895)$ & $f_{\lambda}(1470)$ & err$f_{\lambda}(1470)$ & EW$_{Ly\alpha}$ \\
                                 & &  (deg) & (deg) & &  & (mag)& ($10^{-19}$) & ($10^{-19}$) & ($10^{-19}$) &($10^{-19}$) & ($\AA$) \\
                                 \midrule
                                  & \scriptsize{COSMOS/CANDELS} & & & & & & & & & & \\
                                 \midrule
                          \object{510998698}  & 6772 & 150.082313 & 2.261036  &   4.0651 & 9 & 25.83$\pm$0.12  & 1.98  &  3.23  & 9.04  &  1.49 &  -24.99 \\
                          \object{511002138} & 4913 & 150.122450  & 2.237101   &  4.3600  & 9 & 24.90$\pm$0.06  &  8.25  &  3.26 & 17.08   &  5.63 & -19.18  \\
                          \object{511227001}  & 20282 & 150.062182 & 2.423024   &  3.6350  & 3 & 25.57$\pm$0.11  & 1.79  &  1.39 &   9.26 &  0.48 & -23.84  \\
                          \object{5100998496} & 6868 & 150.205055 & 2.262162   &  3.8979  & 4 & 25.90$\pm$0.15  &  -3.93  & 1.68 &   17.24 & 0.60  &    $\geq  0$   \\
                          \object{5101226001} & 20579 & 150.189291 & 2.427818  &   3.7327  & 3 & 25.71$\pm$0.11 &  1.39   & 2.10  &  11.70 & 0.46 &    $\geq  0$     \\
                          \object{5101226251}  & 20598 & 150.157203 & 2.42786   &   3.9888  & 3 & 25.42$\pm$0.10  &   -4.95 &  1.40 &  17.74  &  0.52 &    $\geq  0$    \\
                          \object{5101233433}*  & 16703 & 150.186851 & 2.379107  &   3.7403  & 4 & 25.19$\pm$0.08 &  5.35  &  1.64  & 18.72 & 0.38 &    $\geq  0$    \\
                          \object{5101233724} & 16397 & 150.177453  & 2.375223   &  4.3862  &   4 & 26.51$\pm$0.20 &  -1.73 &  1.55  & 9.55 &  0.56 & -11.11  \\
                          \object{5101242274}  & 11634 & 150.191107 & 2.317958   &  4.3771  &   4 & 25.95$\pm$0.16 &  -0.34 & 0.80  & 19.61 & 0.67  &$\geq  0$  \\
                                \midrule
                                 & \scriptsize{ECDFS/CANDELS} & & & & & & & & & & \\
                                \midrule
                         \object{530029038}  & 3753 & 53.0792917 & -27.8772595 & 4.4179 & 3 & 26.70$\pm$0.19  & 1.27  &  2.41  &  14.79 & 0.56 &     $\geq  0$      \\
                         \object{530030313}  & 4503 & 53.1132794 & -27.8698754 & 3.5789 & 3 & 26.04$\pm$0.09  & 3.83  &  3.87  &  10.62 & 0.61&   $\geq  0$       \\
                          \object{530030325}  & 4542 & 53.1090745 & -27.8697555 & 3.7519 &   4 & 25.42$\pm$0.06  & -0.34 &  1.41  & 21.25  &  0.43&     $\geq  0$      \\
                          \object{530032655}  & 5955 & 53.0940950  & -27.854974  & 3.7222 & 2 & 25.71$\pm$0.07  & -0.80 &  1.24  &  13.67 &  0.41&    $\geq  0$       \\
                          \object{530036055}  & 8312 & 53.2208728 & -27.8334905 & 4.1608 &  3 &  25.36$\pm$0.05  & -1.29 &  1.21  & 27.90 & 0.85 &     $\geq  0$      \\
                          \object{530037593}  & 9317 & 53.156571  & -27.824343  & 3.5336 &  2 &  25.79$\pm$0.08  & -2.44 &  2.46  &  11.68 &  0.36 &       $\geq  0$   \\                  
                          \object{530047200}  & 17081 & 53.0640040  & -27.765834  & 3.5600    & 2 &  26.06$\pm$0.09  & 6.36  &  14.9   &  16.27 &  1.49 &  $\geq  0$         \\
                         \object{530049753}*  & 18915 &  53.2104941 & -27.7502276 & 3.6055 &  4 & 25.81$\pm$0.08 &  2.92  &  0.93  & 10.68 & 0.20   &    $\geq  0$  \\
                          \object{530049877}  & 18722 & 53.0147863 & -27.7517345  & 3.8245  & 3 & 25.75$\pm$0.08  & 2.53 &   2.20  & 16.14&  0.50  &  -8.43   \\
                          \object{530050023}  & 19009 & 53.2048527 & -27.7494405  & 3.6097 &  4 & 25.07$\pm$0.04  & -3.43 &  2.43 & 21.29 &  0.55 &     $\geq  0$     \\
                          \object{530051970} & 20286 & 53.1989481 & -27.7379129 & 3.7983  &  4 & 24.91$\pm$0.04 &  -1.20   & 1.87 & 26.33  & 0.79 & -5.32  \\                      
                                \midrule
                                 & \scriptsize{ECDFS/GEMS} & & & & & & & & & &  \\
                                \midrule
                         \object{530003871}  & 958 & 53.196500   &  -28.036822  & 3.9022 &  3 & 26.23$\pm$0.12  & 1.61   & 1.25  & 11.81  & 0.49 &  $\geq  0$    \\   
                         \object{530004745}  & 1087 &  53.204005 &  -28.03064   & 3.6447  &  3 & 26.24$\pm$0.11 &  -3.48 &  2.8  & 12.27  & 0.66  &  -18.61   \\
                          \object{530008598}  & 432 & 53.004328 &  -28.006944  & 4.0409 &  4 &  25.04$\pm$0.04 &  2.73  &  2.32 & 40.80  &  0.91  & -15.72  \\
                          \object{530010169}  & 624 & 53.209205 &  -27.996984  & 3.9443 &  3 & 26.65$\pm$ 0.18 &  -1.41 &  1.13  &  12.59 & 0.52 &       $\geq  0$     \\
                          \object{530013910}  & 233 & 53.222218  & -27.973808  & 3.9560  &  4 & 24.59$\pm$0.03  & 4.21  &  1.52  &  48.19 &   0.82& -2.76   \\          
                         \object{530046805} & 164 &  52.98375   & -27.768377  & 3.5748 &  4 & 25.03$\pm$0.04   & 0.35  &  1.55  &  20.63 & 0.34&    $\geq  0$     \\
                         \object{530063568}  & 629 & 53.198749  & -27.666776  & 4.1005 & 4 &  26.10$\pm$0.11 &  4.18   & 2.32 &  13.99  & 0.75&    $\geq  0$       \\
                         \object{530070083} & 22 & 53.048402  & -27.626923 &  3.6377  & 3 & 24.68$\pm$0.03 &  -5.93  & 2.24 &   26.74 &   0.76&   $\geq  0$     \\
                         \object{530074267}  & 322 & 53.135104  & -27.601156  & 3.5125 &  3 & 25.27$\pm$0.05  &  -1.22  & 3.17 &  13.86  & 0.41 &     $\geq  0$     \\
                          \object{530075924}*  & 1256 & 53.060854  & -27.590558 &  3.5707 & 3 & 25.24$\pm$0.05 &  5.68   & 1.38 &  12.25  &  0.36 & -65.00 \\
                         \object{530076805}  & 1402 & 53.247191 &  -27.584018 &  3.9602  & 2 & 26.64$\pm$0.18 &  -3.36  & 2.77 &  16.30  &  1.04& -1.46 \\
                         \object{535003653}  & 852 & 53.080489  & -28.037964  & 4.2093  &  3 & 26.70$\pm$0.18 &  1.95   & 0.88 & 4.26  & 0.45 & -160.39\\
                         \object{535016440} & 1576 & 53.051566  & -27.957000    &  4.3215  &  3 & 27.93$\pm$0.58 &  -0.60   & 0.88 &   5.41 &  0.37&     $\geq  0$     \\
        
                        \bottomrule[1.5pt]
                        \end{tabular} 
                        \vspace{0.1cm}
                        \caption{Sources in the clean sample. (1) VUDS identification number; (2) CANDELS or GEMS
                        identification number; (3)(4) Right ascension and declination; (5) Spectroscopic redshift; (6)
                        Redshift quality flag \citep[see][for details]{lefevreVIMOS}; (7) V magnitude and respective error; (8)(9) Mean flux in the wavelength range $880-910 \AA$ and respective error in unit of $10^{-19} erg\, s^{-1}\, cm^{-2}\, \AA ^{-1}$; (10)(11) Medium
                        flux in the wavelength range $1420-1520 \AA$ and respective error in units of $10^{-19} erg\, s^{-1}\, cm^{-2}\, \AA ^{-1}$;
                        (12) Ly$\alpha$ equivalent width of the sources that present Ly$\alpha$ in emission. We have indicated the Ly$\alpha$ in absorption with $\geq  0$. We have also marked with * the \emph{possible leakers} in Section \ref{sub:fluxratio}.}
                        \label{tab:cleansample}
                }
                \end{table*}
        
                \section{Measuring the LyC signal: the escape fraction of ionising photons}
                \label{sec:fesc}
                
                The escape fraction  is the fraction of Lyman continuum photons produced by massive OB-type stars in star-forming galaxies that escape from the galaxy into the IGM without being absorbed, relative to the total number produced \citep{fesc,fesc2}, and references therein). This quantity, known as 
                the \emph{absolute escape fraction}, $f_{esc}$, requires knowledge of the intrinsic number of
                ionising photons produced by the galaxy to determine it. However, the intrinsic spectral energy distribution (SED) of a galaxy is not known a priori, especially in the rest-frame far UV where dust reddening could be severe.

                 A related quantity, more used in observational studies, is the \emph{relative escape fraction},  which is defined as \citep{steidel,siana}:
                \begin{equation}
                \label{eq:fesc}
                f_{esc}^{rel}(LyC)=\frac{L_{\nu}(1470)/L_{\nu}(895)}{f_{\nu}(1470)/f_{\nu}(895)\cdot e^{-\tau_{IGM,z}}}
                ,\end{equation}
                where $L_{\nu}(1470)/L_{\nu}(895)$ is the ratio of the intrinsic luminosities
                at 1470 and 895 $\AA$ rest frame and $f_{\nu}(1470)/f_{\nu}(895)$ is the ratio of the
                observed flux densities at the same wavelengths. $e^{-\tau_{IGM,z}}$ is the \emph{transmissivity}, a quantity that allows us to take into account  the photoelectric absorption of photons with $\lambda \leq 912 \AA$ by the IGM.
                $f_{esc}^{rel}(LyC)$ is a useful  observational quantity  because it can be defined with respect
                to the  UV continuum luminosity, which is accessible to direct observations. For star-forming galaxies in fact, at $z\sim 4$,
                the continuum flux at $\lambda \sim 1470\AA$ used in equation \ref{eq:fesc} can be  determined from observations.
                It is also possible to determine $f_{esc}$ from $f_{esc}^{rel}$ if the degree of dust extinction in the UV is known: $f_{esc}$ is defined indeed as the relative escape fraction multiplied by the extinction of the dust. In the absence of dust, the two quantities are equivalent.
                
                In the following subsections, we describe our procedures to derive the quantities in Eq. \ref{eq:fesc} to estimate the relative escape fraction of our clean sample of galaxies.
                \subsection{Intrinsic L$_{\nu}$(1470)/L$_{\nu}$(895) ratio and IGM  transmissivity}
                \label{sub:transmittivity}
                The intrinsic luminosity ratio $L_{\nu}(1470)/L_{\nu}(895)$
                depends on the physical properties of the galaxies, such as the
                mean stellar ages, metallicities, stellar initial mass functions (IMFs),
                and star formation histories (SFHs) as also shown in \cite{lucia} (see Table 3); for typical star-forming
                galaxies at $z \sim 3$,  it varies between 1.7 and 7.1 for an age between 1 Myr and 0.2 Gyr,
                adopting the \cite{bruzual} library \citep{grazian16}.
                For easier comparison with earlier studies \citep{steidel,grazian16}, we adopted a value of $L_{\nu}(1470)/L_{\nu}(895)=3,$ which corresponds to young star-forming galaxies of $\sim 10$ Myr assuming a constant Bruzual \& Charlot SFH and a Chabrier IMF \citep{chabrier03}. We checked the best fit physical parameters of the galaxies in the
                CANDELS fields \citep{santini15} in our sample, and for most models their age is
                actually closer to $100-200 Myr$. Therefore, a more appropriate value for
                the intrinsic $L_{\nu}(1470)/L_{\nu}(895)$ would be 5-6  \citep{lucia}.
                However we note that this is simply a multiplicative factor in the
                evaluation of the relative escape fraction, and it is therefore possible to re-scale $f_{esc}^{rel}(LyC)$ with other values of $L_{\nu}(1470)/L_{\nu}(895)$ if needed.
                
                We computed the IGM  transmissivity following the analytical prescription given by \cite{inoue:transmissivity}. They provide a set of analytic functions describing the mean intergalactic attenuation curve for objects at $z>0.5,$ simulating, for different lines of sight, a large number of absorbers, and assuming a Poisson probability distribution for a LyC photon to encounter one of them. The Monte Carlo simulations are based on an empirical distribution function of intergalactic absorbers that they derived from the latest observational statistics of the Ly$\alpha$ forest (LAF), Lyman limit system (LLSs) and damped Ly$\alpha$ systems (DLSs), and is verified to be consistent with the mean Ly$\alpha$ transmission and the mean free path of ionising photons at those redshifts \citep{worseck14}.  We refer to \cite{inoue:transmissivity} for understanding the limitations and uncertainty related to the application of this prescription. We highlight here that there is  a large 
scatter in the IGM transmission around the mean at each given redshift, as shown, for example, in Fig. 2 of \cite{vanzella15}.

                \subsection{Measuring  $f_{\lambda}(895)$ and $f_{\lambda}(1470)$ }
                \label{sub:fluxratio}
                \begin{figure}
                \centering
                \includegraphics[width=\hsize]{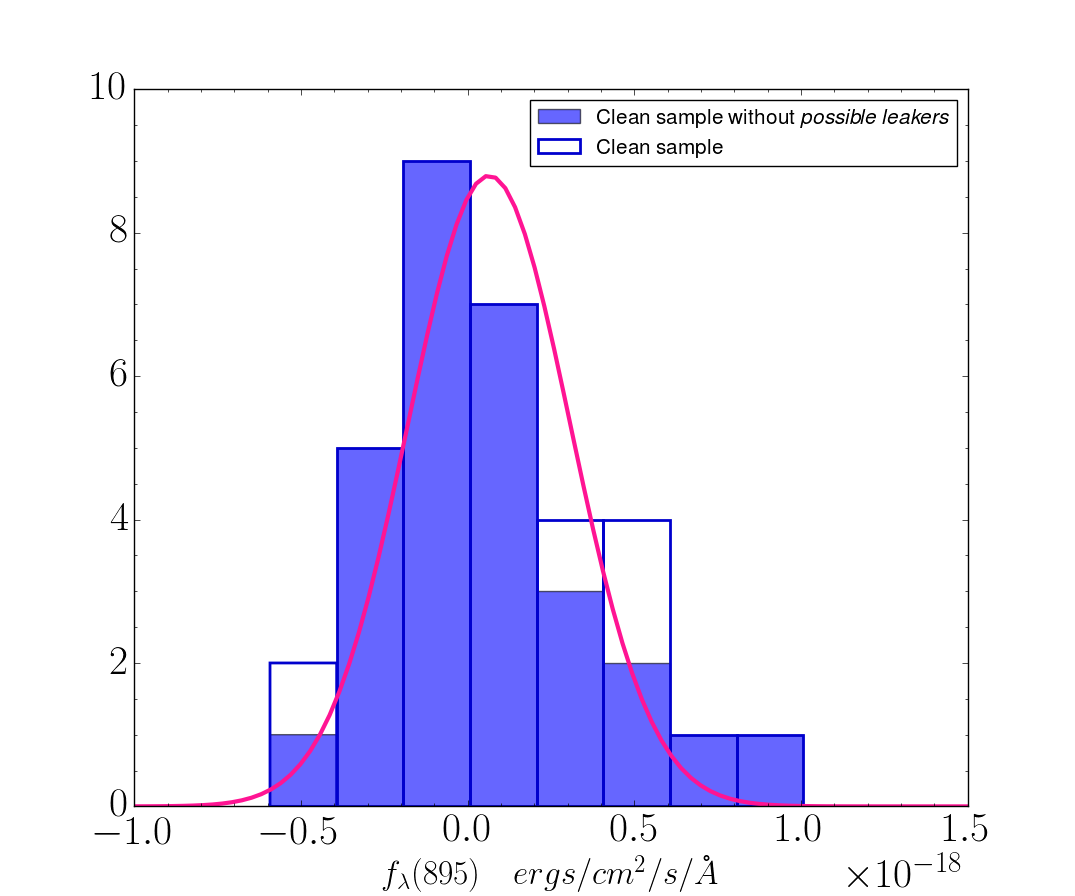}
                \caption{Distribution of the fluxes measured as the mean fluxes in the LyC range of each
                spectrum ($880-910 \AA$) for the sources that have LyC flux within three times their standard deviation (filled histogram) and for the clean sample (empty histogram). The Gaussian fit of the distribution without the \emph{possible leakers} is shown in magenta. }
                \label{fig:histoF}
                \end{figure}
                
                For each source in the clean sample, we evaluated from the spectrum the UV flux as the mean value in the
                wavelength interval $1420-1520 \, \AA$, and the LyC flux, or a limit, in the range $880-910\, \AA$ ($f_{\lambda}(1470)$ and
                $f_{\lambda}(895)$). We chose the $1420-1520 \, \AA$ interval for the UV continuum because in this range, no significant spectral features are present. The excellent relative flux calibration adds only a negligible error when computing the $f_{\lambda}(895)/f_{\lambda}(1470)$ ratio, and this is therefore ignored in the following.
                
                We report the individual values  in Table \ref{tab:cleansample} with the respective individual statistical errors  (${\rm err}f_{\lambda}(1470)$ and ${\rm err}f_{\lambda}(895)$).
                As a  first approximation, we considered, as non-LyC emitters, all the sources
                with $f_{\lambda}(895)$ flux consistent
                with zero within three times their individual statistical error,
                 and consider possible LyC
                emitters, as the sources that have values exceeding 3$\times {\rm err}f_{\lambda}(895)$ (hereafter referred to as \emph{possible leakers}). Three of these have a positive $f_{\lambda}(895)$ and one has a negative $f_{\lambda}(895)$.
                However, the individual errors  do  not  take into
                account all the possible systematic uncertainties associated, for example, with background
                subtraction (which change from slit to slit), with the presence of
                scattered light, or with other quantities that are related to the instrument setup for the
                masks used. Considering that the objects in our samples were included in many different masks and observed under very  different conditions,  we decided to assess the reality of the possible detections by re-evaluating the global error from the distribution of the $f_{\lambda}(895)$ values for the remaining (undetected) objects. 
                
                Excluding the possible leakers, we computed the distribution
                of   the LyC fluxes of the non-LyC emitting sources, as shown  in Fig.
                \ref{fig:histoF}; in
                magenta, we show  the Gaussian fit to the distribution that was evaluated after applying a sigma clipping with a limit at $2\sigma$. This  is
                characterised by a
                mean value of $f_{\lambda}(LyC) = 0.86 \cdot 10^{- 20}$ erg  s$^{-1}$
                cm$^{-2}$ \AA$^{-1}$ and a standard deviation of $\sigma_s= 2.39 \cdot
                10^{-19}$ erg  s$^{-1}$ cm$^{-2}$ \AA$^{-1}$. The mean value is therefore
                basically consistent with zero (as expected).
                On the other hand, the statistical  error $\sigma_s= 2.39 \cdot
                10^{-19}\, erg  s^{-1}$ cm$^{-2}$ \AA$^{-1}$  is similar or somewhat larger (by 50\%) than most
                of the individual errors -  $f_{\lambda}(895)$ in Table \ref{tab:cleansample} (${\rm err}f_{\lambda}(895)$.   
                We thus  assume this value as the  mean  error in $f_{\lambda}(895)$ for our whole sample; with this assumption,  we do not have sources that still show a detection of LyC above 3$\sigma$. All the galaxies in the clean sample therefore have LyC fluxes consistent with zero at 3$\sigma$ level.

                \section{Spectral stack}
                \label{sec:stack}
                Since we do not have any solid LyC detection in the individual galaxies, we produced and analysed a stacked spectrum of all sources belonging to  the clean sample to increase the sensitivity of our measurement.
                
\begin{figure*}
                \includegraphics[width=\textwidth]{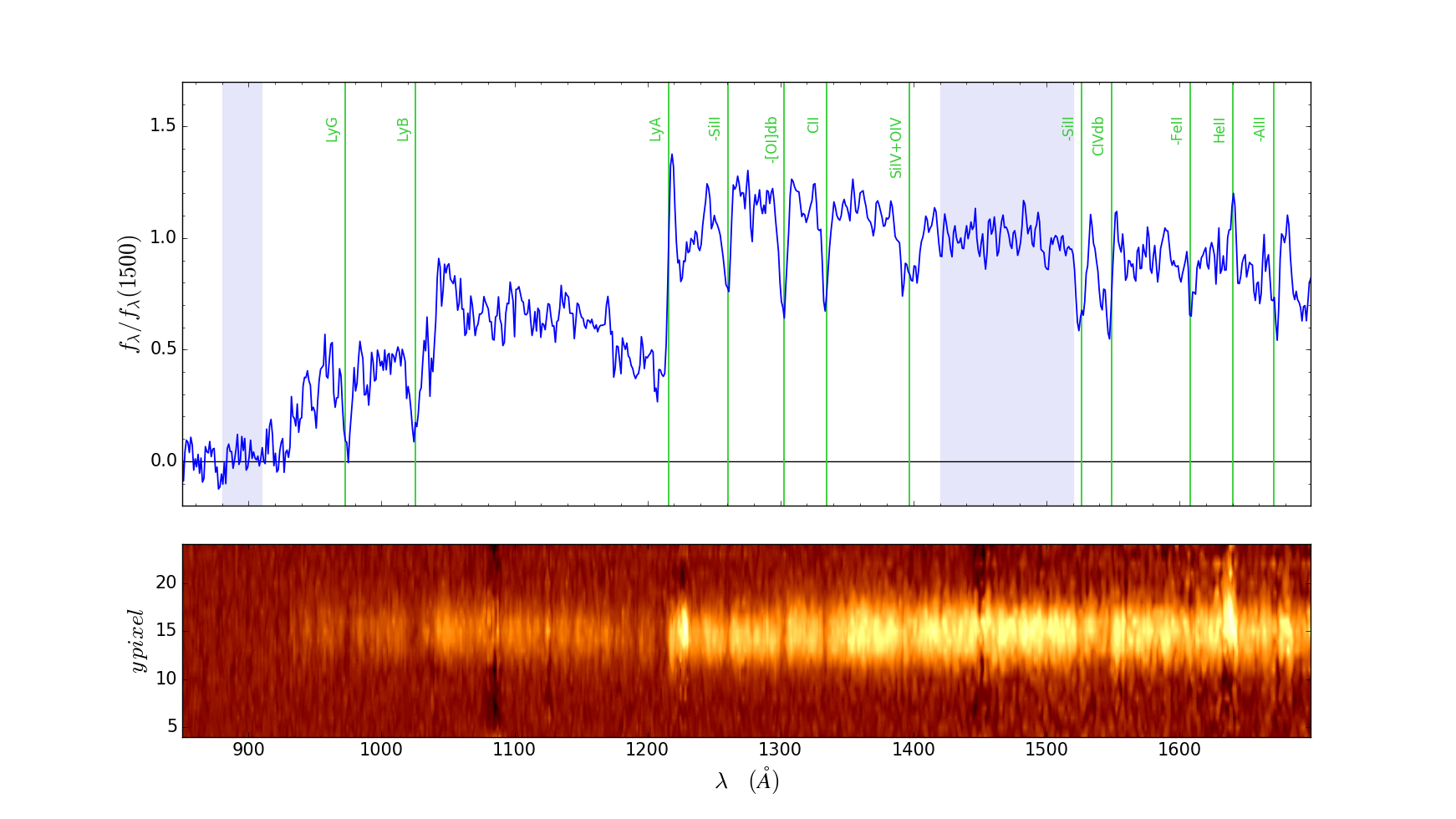}
                \caption{One-dimensional and two-dimensional spectral stack of the 33 galaxies in the clean sample. $f_{\lambda}/f_{\lambda}(1470)$ corresponds to the flux normalised at its value at $1470\AA$. The lavender vertical bands represent the spectral region in which we extracted the LyC flux and the UV flux.}
                \label{fig:stack}
                \end{figure*}

                To produce  the one-dimensional spectral stack, we first shifted each
                spectrum to its rest-frame using the spectroscopic redshift from VUDS and normalised
                it using its mean value in the wavelength range $1420-1520 \, \AA$ where no particular features
                are present. To take into account the noise of each spectrum during the stacking procedure, we computed the stack as a weighted average of the spectra in the clean sample using the errors related to the flux density ratio $\frac{f_{\lambda}(895)}{f_{\lambda}(1470)}$ 
                 as weights (these had been previously evaluated). We obtained the spectral stack shown in Fig. \ref{fig:stack}. 
                 Note that in this procedure, we do not use precise systemic redshifts, which should be computed only from the inter-stellar absorption lines or nebular emission lines, and which in many cases cannot be evaluated, but we rely on the VUDS official redshifts \citep[see][for details on the redshift evaluation]{lefevreVIMOS}. 
                The presence of sharp absorption features in the stack indicates that we can safely use the spectroscopic redshifts in our procedure, since our goal is to analyse the LyC region that covers a broad wavelength range.
                
                We also computed the two-dimensional spectral stack, which is presented  in Fig. \ref{fig:stack}. After shifting each spectrum to the same spatial position (along the slit length) and to its wavelength rest frame, we applied the same weighted average used for the one-dimensional stack. The extension of the obtained spectrum is due to the fact that each spectrum has a different spatial width due to different object sizes as projected onto the slit. 
                
                \section{Results}
                \label{sec:res}
                We now derive the average relative escape fraction of the clean sample.  
                The first quantity that is needed is an approximation of the $\frac{f_{\nu}(895)}{f_{\nu}(1470)}$ ratio of the clean sample. We evaluate it following two different approaches. The first involves  directly measuring it from the spectral stack (Section \ref{sec:stack}), that is, from the average signal in the wavelength range $880-910\, \AA$. The flux in the spectral stack in Fig. \ref{fig:stack} is normalised at the flux value in the range $1420-1520 \, \AA$. We obtain a value of $\frac{f_{\nu}(895)}{f_{\nu}(1470)}=0.008\pm 0.004$.  2) The second method consists of evaluating  the flux density ratio for each source in the clean sample and then computing  a weighted average using the errors of the ratios that were derived above  as weights. With this method, we find $\frac{f_{\nu}(895)}{f_{\nu}(1470)}=0.008\pm 0.005$.  Therefore the two methods give almost identical values.
                
                 The value of the transmissivity depends on the redshift of the galaxies considered. We proceeded with two different approaches for this quantity as well. 
The first consists of  evaluating the transmissivity for each object in the sample and then averaging them. We call this the \emph{mean transmissivity} of the sample and we consider this method as the one with a more physical meaning, since it takes into account the uneven redshift distribution of our sample, which is skewed towards the lower redshifts. The second 
method  involves directly evaluating the transmissivity at  the median redshift  of the clean sample, $z_{MED}=3.81$. We obtain values of $e^{-\tau_{IGM,z}}=0.27$ and $e^{-\tau_{IGM,z}}=0.29,$ respectively.

                Using different combinations of these values, we computed the relative escape fraction of the clean sample. 
                The values that we found are listed in Table \ref{tab:fescsenza}. The errors are evaluated propagating the error on the flux density ratio to Eq. \ref{eq:fesc}.
From the stacked spectrum, we obtain tentative $\geq 2\sigma$ detections for both the evaluation methods, whereas using the average of individual signals, the overall errors are larger.  In all cases, the values are consistent with a null relative escape fraction within $2\sigma$.
                
                        \begin{table*}
                        \centering
                                \begin{tabular}{l|c| c}
                                        \toprule
                                        \toprule
                
                                        &Mean transmissivity  &Transmissivity from the median redshift \\
                                        \midrule
        Flux ratio from the stack & $f_{esc}^{rel}=0.09\pm 0.04$ & $f_{esc}^{rel}=0.08\pm 0.04$ \\ 
        \hline
                Flux ratio from the average  & $f_{esc}^{rel}=0.09\pm 0.06$& $f_{esc}^{rel}=0.08 \pm 0.05$\\

                                        \bottomrule[1.5pt]
                                \end{tabular} 
                                \vspace{0.1cm}
                                \caption{Relative escape fraction of the clean sample 
                                for the different combinations of transmissivities and $\frac{f_{\nu}(895)}{f_{\nu}(1470)}$ ratios that we found with the different methods explained in Sec. \ref{sec:res}.}
                                \label{tab:fescsenza}
                        \end{table*}

                \section{Discussion and future prospects}
                 Several authors have discussed the possible correlation between the escape of LyC photons and the escape of Ly$\alpha$ photons. Recently,  \cite{dijkstra16}   used  a suite of 2500 Ly$\alpha$ Monte-Carlo radiative transfer simulations to show that galaxies with a low $f_{esc}^{Ly\alpha}$ consistently have a low $f_{esc}^{LyC}$. It is believed that very compact galaxies with a strong Ly$\alpha$ in emission are the most plausible  candidates to show LyC leakage. \cite{slavo} found that their LyC-detected sample showed a significantly stronger emission of Ly$\alpha$ photons on average, with respect to the non-detected sample, although their sample was prone to low redshift contamination since it lacked high-resolution HST imaging. Finally, an empirical correlation between EW(Ly$\alpha$) and $f_{esc}(LyC)$ has also been shown by \cite{verhamme16} for the small number of LyC confirmed sources, including both local galaxies and Ion2 \citep{vanzella15}.

We therefore investigated whether or not the limits on the escape fraction of the  individual sources in our sample have any correlation to the presence of Ly$\alpha$.
                For each spectrum that shows Ly$\alpha$ in emission, we measured the equivalent width (EW) of the line
                using IRAF.  We report the values of the EW of the sources in our sample in Table \ref{tab:cleansample} where we have only indicated $EW(Ly\alpha)\geq 0$ for the spectra that show the line in absorption. We compared the EW derived with this method to those derived by \cite{cassata:EW} and found, in general, a very good agreement.
                
\begin{figure}[htp]
                  \centering
                  \subfigure{\includegraphics[width=0.5\textwidth]{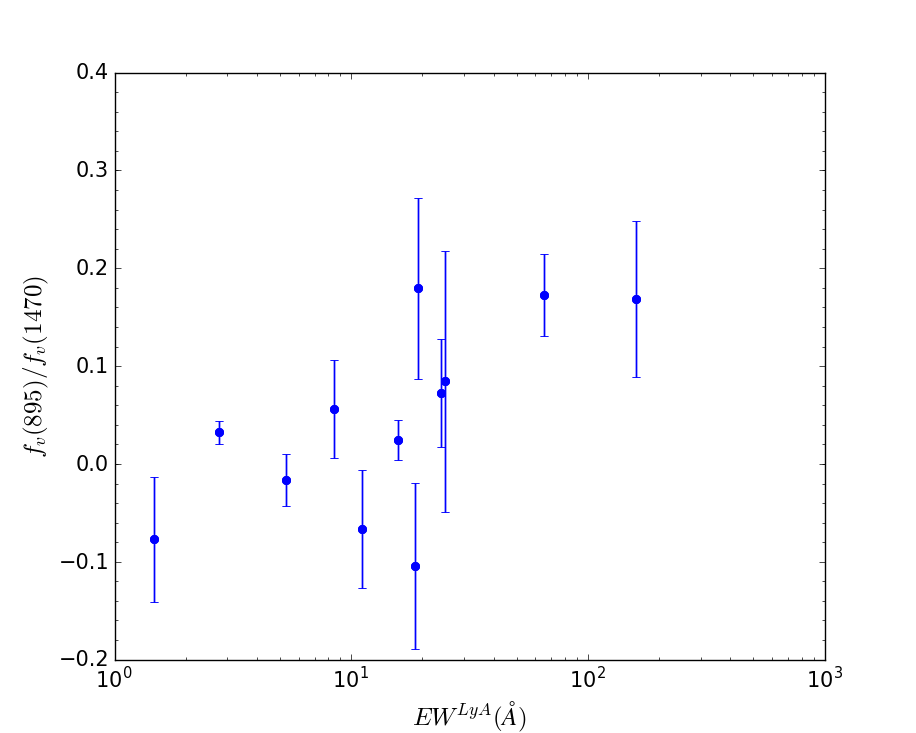}}
                  \caption{Ratio between the mean fluxes at $895\AA$ and $1470\AA$ in units of $ergs\, s^{-1}cm^{-2}Hz^{-1}$ as a function of the absolute value of the rest frame equivalent width of the Ly$\alpha$ for the 12 galaxies  that have Ly$\alpha$ in emission.}
                  \label{fig:fratio_EW}
                \end{figure}

\begin{figure}
                \centering
                \includegraphics[width=0.45\textwidth]{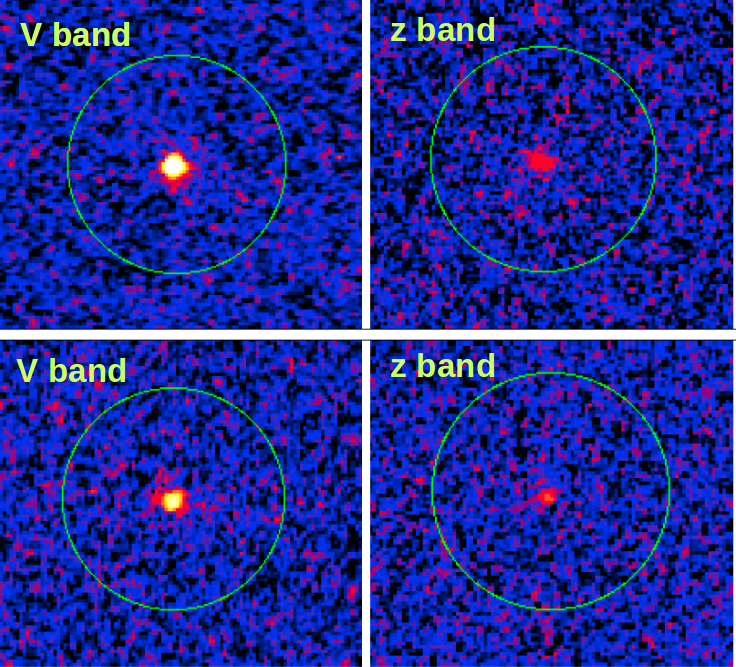}
                \caption{GEMS images in the V and z band of the two sources: \object{530075924} (\emph{Upper panel}) and \object{535003653} (\emph{Bottom panel}). The circular regions have 1 arcsec radius.}
                \label{fig:compact}
                \end{figure}

                In Fig. \ref{fig:fratio_EW}, we show the flux density ratio as a function of the absolute value of the rest frame equivalent width of the Ly$\alpha$, for the sources that have the line in emission.  
                The errors on the flux density ratios have been evaluated using the individual errors of each object.  Despite the fact that we have few Ly$\alpha$ emitters in our sample, from the figure, we see  indications of  a possible trend of the flux density ratio as a function of Ly$\alpha$ EW, in the sense that, if Ly$\alpha$ is in emission, the ratio increases with increasing EW. This trend could therefore indicate that the mechanisms that drive the escape of Ly$\alpha$ photons also facilitate the escape of LyC radiation. The same tentative trend is found by \cite{micheva15} with a sample of 18 LAEs at $z\geq 3.06$.
                 In particular, we point out that the only two sources in the clean sample that show a very strong
Ly$\alpha$ emission, have a high flux density ratio with
respect to the average of the other sources (they are
the two upper right points in  Fig. \ref{fig:fratio_EW}).
In addition, both have a very  compact morphology, as shown in Fig. \ref{fig:compact}, so
they would be good LyC emitter candidates according to \cite{izotov16}, who show that  selecting 
 compact star-forming galaxies with high $[OIII]\lambda5007/[OII]\lambda3727$
ratios appears to very efficiently pick up sources with escaping Lyman
continuum radiation.
In particular,  one of the our two sources, namely VUDS ID \object{530075924},  was  identified as a
possible emitter in the initial screening of the sample; when considering its  individual statistical errors, the  LyC flux was
detected with a  $>4\sigma$ significance, and it was therefore
included in the \emph{possible leakers} sample (see Section \ref{sub:fluxratio}).
However, after the re-evaluation of the error from the sample distribution,
the significance of the detection dropped to $2.5\sigma$ and  it
was therefore considered consistent with a non-emitter.
Given the high flux density ratio, the  large  Ly$\alpha$ EW and the compact morphology, 
this galaxy could, in fact, be considered our best candidate LyC leaker. In addition, this object also has the same properties of the galaxies described in Amorin et al. (in prep), which are selected to be extremely young (metal-poor) dwarf galaxies, typical features of LyC leakers at low redshift.
Taking the LyC flux at face value and the mean transmissivity at that redshift, this source would have a relative escape fraction $>100 \%$ with a very large uncertainty. This extremely high value could only be considered physical in the specific cases in which the galaxy resides in a region with a clearer line-of sight (thus with higher transmissivity) compared to the average, or if the galaxy had a smaller intrinsic  L$_{\nu}$(1470)/L$_{\nu}$(895) than the standard value assumed. Alternatively, it could be characterised by a different dust geometry, as in the case of runaway massive stars that emit LyC radiation near the border of the galaxy where the coverage of the dust is less effective \citep{conroy12}.
We plan to further investigate the nature of this source, however, we  note that to eventually confirm its nature as a LyC leaker
would require a significant amount of integration time with currently existing instrumentation, given that the VUDS spectrum is already the result of 14 hours integration on the VLT.

We finally  checked if the limits on the escape fraction of individual sources have any correlation to the UV magnitude but no trend is observed in agreement with what has previously been found; by \cite{lucia}, for example.

        In conclusion, in  this study, we do not find any solid individual
        detection of LyC emission, although we have a possible interesting candidate, discussed
        above.  We obtain an average relative
        escape fraction of the clean sample of $f_{esc}^{rel}=0.09\pm0.04$ by stacking all the individual spectra. This tentative detection is consistent with 
 what we found in our previous study \citep{lucia} based on a similar clean sample, that is, not effected by low-redshift contamination, and studied through narrow band photometry, for which we reported a $f_{esc}^{rel}<12\%$. This value is also consistent with what has found previously by
 \cite{grazian16} and \cite{boutsia} at a similar redshift. This is also in agreement with \cite{vanzella10}. They used ultra-deep ultraviolet VLT/VIMOS intermediate-band and VLT/FORS1 narrow-band imaging in the GOODS-S field to derive limits on the distribution of the absolute escape fraction for LBGs in the redshift interval $3.4-4.5$. They found a median $f_{esc}$ lower than $\sim 6\%$ with an $84\%$ percentile limit not larger than $20\%$ that translates to a median $f_{esc}^{rel}$ lower than $\sim 11\%$.
                Our result is instead not in agreement with \cite{micheva15} and \cite{smith16}. \cite{micheva15} analysed samples of 18 LAEs and 7 LBGs at $z\ge 3.06$ obtained from the SSA22 field with Subaru/Suprime-cam. They found values for the relative escape fraction of $\sim 30\%$ for the LAEs sample and $\sim 20\%$ for the LBGs sample. These high values could however be due, in part, to foreground contamination, since many LyC emitter candidates in their samples show a spatial offset between the rest-frame UV and LyC emissions. Contamination from lower redshift interlopers could also be present in \cite{smith16} who presented observations of escaping LyC radiation from 50 galaxies in different redshift bins from $z\sim 2.3$ to $z\sim 5.8$ in the Early Release Science (ERS) field. They found $f_{esc}^{rel}=19.8^{+39.2}_{-10.6}$ for galaxies at $z\sim3.5$;
at $z\sim 5$ however the relative escape fraction exceeds $100 \%$, indicating a possible contamination effect.
                
                We plan to  apply the same method   to another upcoming  spectroscopic survey, VANDELS (\href{url}{http://vandels.inaf.it/}), that will give us, in one year, spectra of high-redshift galaxies  with unprecedented observation time of up to 80 hours, also with VIMOS.
                With these spectra, we might be able to  detect LyC emitters or set much stronger constraints on the average escape fraction of high-redshift galaxies.  In particular, by using the target catalogues in the CDFS and UDS fields and the relative photometric redshifts,  we have estimated that we will be able to study between 100-150 LBGs and star-forming galaxies at $4.4\leqslant z_{phot}\leqslant 5.0$ where the LyC range is still included in the VANDELS spectra. 
                Comparing our predictions with the sensitivity of the instrumentation, we have estimated that we will be able to detect  the LyC in individual LBGs  if $f_{esc}^{rel}>0.05-0.2$ and average signals 
                of $f_{esc}^{rel}\sim 0.02-0.05$ from the stacks of 30-50 objects samples, thus setting much tighter constraints on the properties of high-redshift galaxies.
        \begin{acknowledgements}
We thank the ESO staff for their continuous support for the VUDS survey,
particularly the Paranal staff conducting the observations and Marina Rejkuba and
the ESO user support group in Garching.
This work is supported by funding from the European Research Council Advanced Grant
ERC--2010--AdG--268107--EARLY and by INAF Grants PRIN 2010, PRIN 2012 and PICS 2013.

AC, OC, MT and VS acknowledge the grant MIUR PRIN 2010--2011.  
This work is based on data products made available at the CESAM data center,
Laboratoire d'Astrophysique de Marseille. 
 R.A. acknowledges support from the ERC Advanced Grant 695671 ‘QUENCH’.
        \end{acknowledgements}
   \bibliographystyle{aa} 
   \bibliography{biblio}
   
   \begin{appendix}
   \section{The importance of the cleaning procedure}
\begin{figure*}
   \centering
   \includegraphics[width=0.8\textwidth]{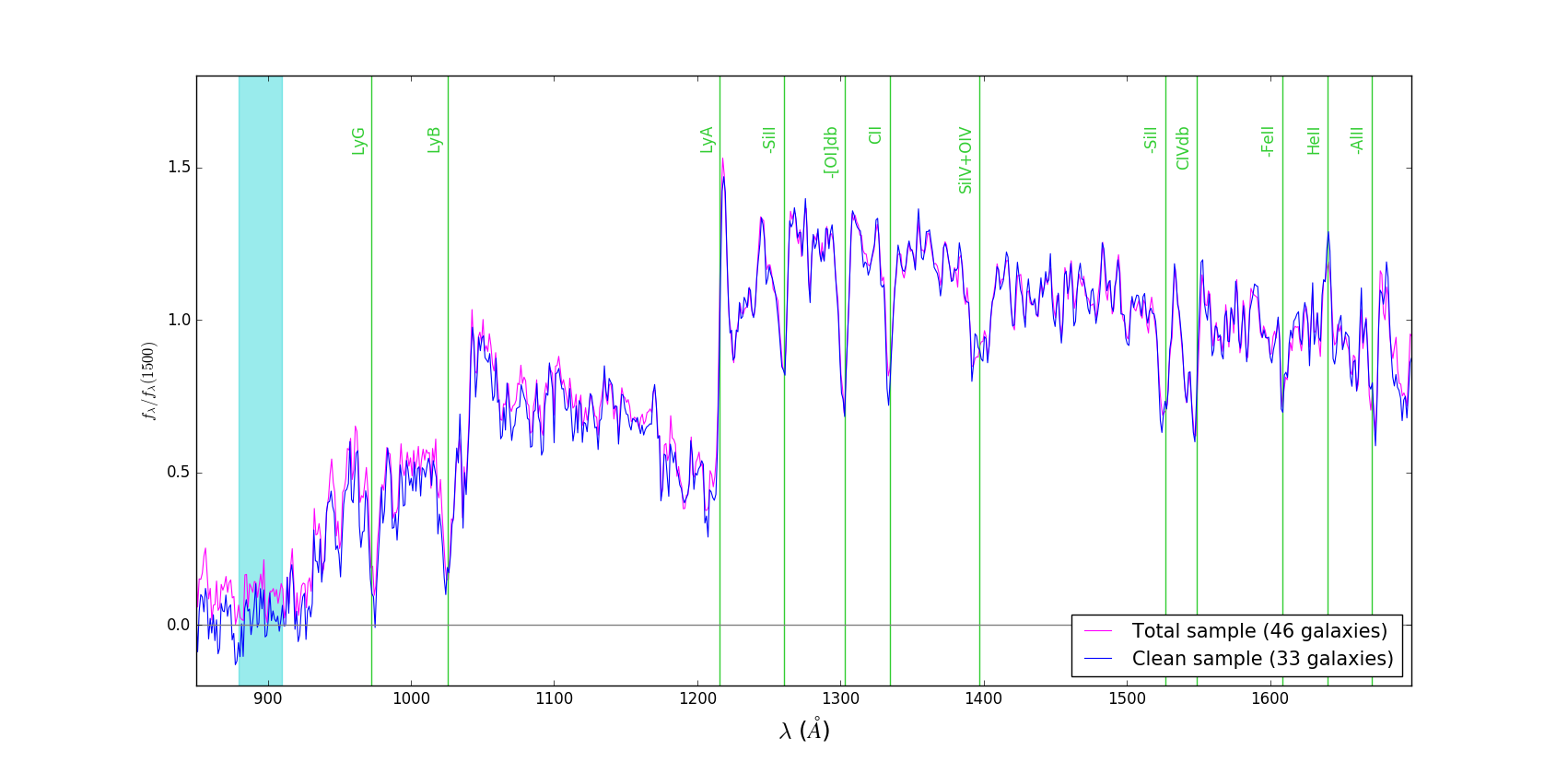}
   \caption{One-dimensional stack of the clean sample (blue line) compared with the stack of the total sample before the application of the cleaning procedure (magenta line).}
   \label{fig:stacks}
   \end{figure*}

   We show in Fig. \ref{fig:stacks} the comparison between the stack of the clean sample (blue spectrum) and the stack of the total sample (magenta spectrum) in order to point out the importance of the cleaning procedure in these kind of studies. The flux of the total sample is indeed a factor of 5 greater than the flux of the clean sample in the LyC region (turquoise vertical band in Fig. \ref{fig:stacks}) whereas they have the same values in the UV part of the spectrum. This exceeding flux is evidence of the contamination from lower redshift interlopers that we avoided thanks to the cleaning procedure.
   \end{appendix}
   
\end{document}